\documentclass{ws-procs9x6}

\begin{document}

\title{Transverse Spin Studies with STAR at RHIC}

\author{C.A. Gagliardi$^1$,
for the STAR Collaboration\footnote{http://www.star.bnl.gov}}

\address{$^1$\,Cyclotron Institute, Texas A\&M University \\
College Station, TX, 77843, USA\\ 
E-mail: cggroup@comp.tamu.edu}

\maketitle

\abstracts{
STAR has observed sizable transverse single-spin asymmetries for inclusive $\pi^0$ production at forward rapidity in p+p collisions at $\sqrt{s}=200$ GeV.  These asymmetries may arise from either the Sivers or Collins effect.  Studies are underway during the current RHIC run to elucidate the dynamics that underlie these single-spin asymmetries.  Additional measurements are underway to search for the Sivers effect in mid-rapidity di-jet production.}

\section{Introduction}

During RHIC Run 2, the STAR Collaboration observed sizable transverse single-spin asymmetries, $A_N$, for $\pi^0$ production at large $x_F$ in $\sqrt{s}=200$ GeV p+p collisions,\cite{pi0_PRL} similar to previous measurements at $\sqrt{s}=20$ GeV.\cite{FNAL_pi0}  The previous asymmetries were observed in a regime where pQCD significantly underpredicts the inclusive cross section.\cite{Bour04}  In contrast, pQCD gives a good description of the large-$x_F$ inclusive $\pi^0$ cross section at $\sqrt{s}=200$ GeV.\cite{pi0_PRL,dAu_PRL}  Nonetheless, the observed asymmetries are reproduced qualitatively by pQCD models fitted to the data at 20 GeV, then extrapolated to 200 GeV.\cite{PRL_theory}  The Sivers model introduces a spin and $k_T$-dependence to the parton distribution functions.  A second model convolutes transversity with a spin-dependent Collins fragmentation function.  Other models include twist-3 parton correlations in the initial or final state, which are closely related to the Sivers and Collins effects.\cite{DIS06}

\begin{figure}[t]
  \begin{minipage}{0.49\textwidth}
    {\epsfxsize=\textwidth\epsfbox{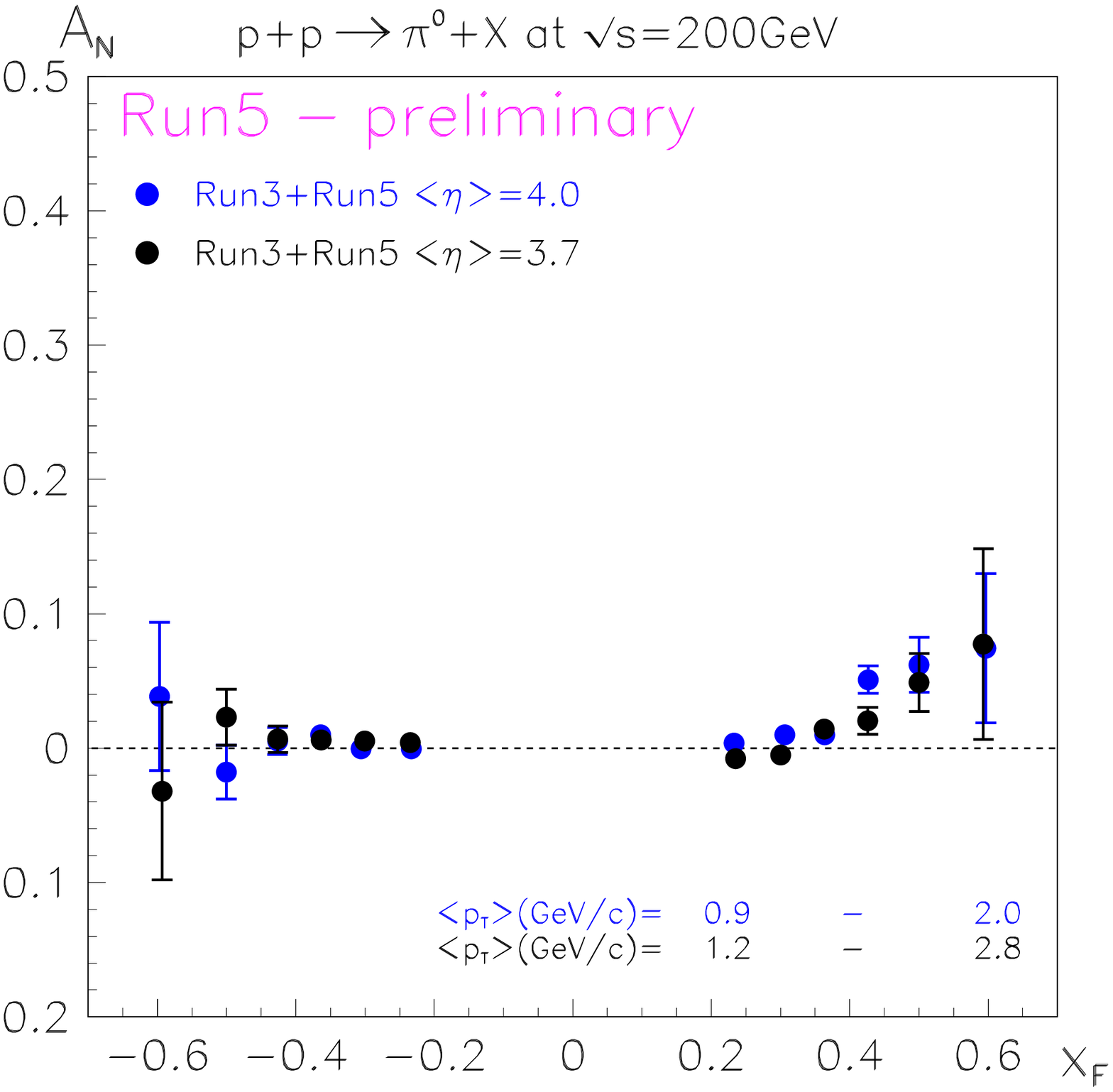}}
  \end{minipage}
  \hfill
  \begin{minipage}{0.49\textwidth}
    {\epsfxsize=\textwidth\epsfbox{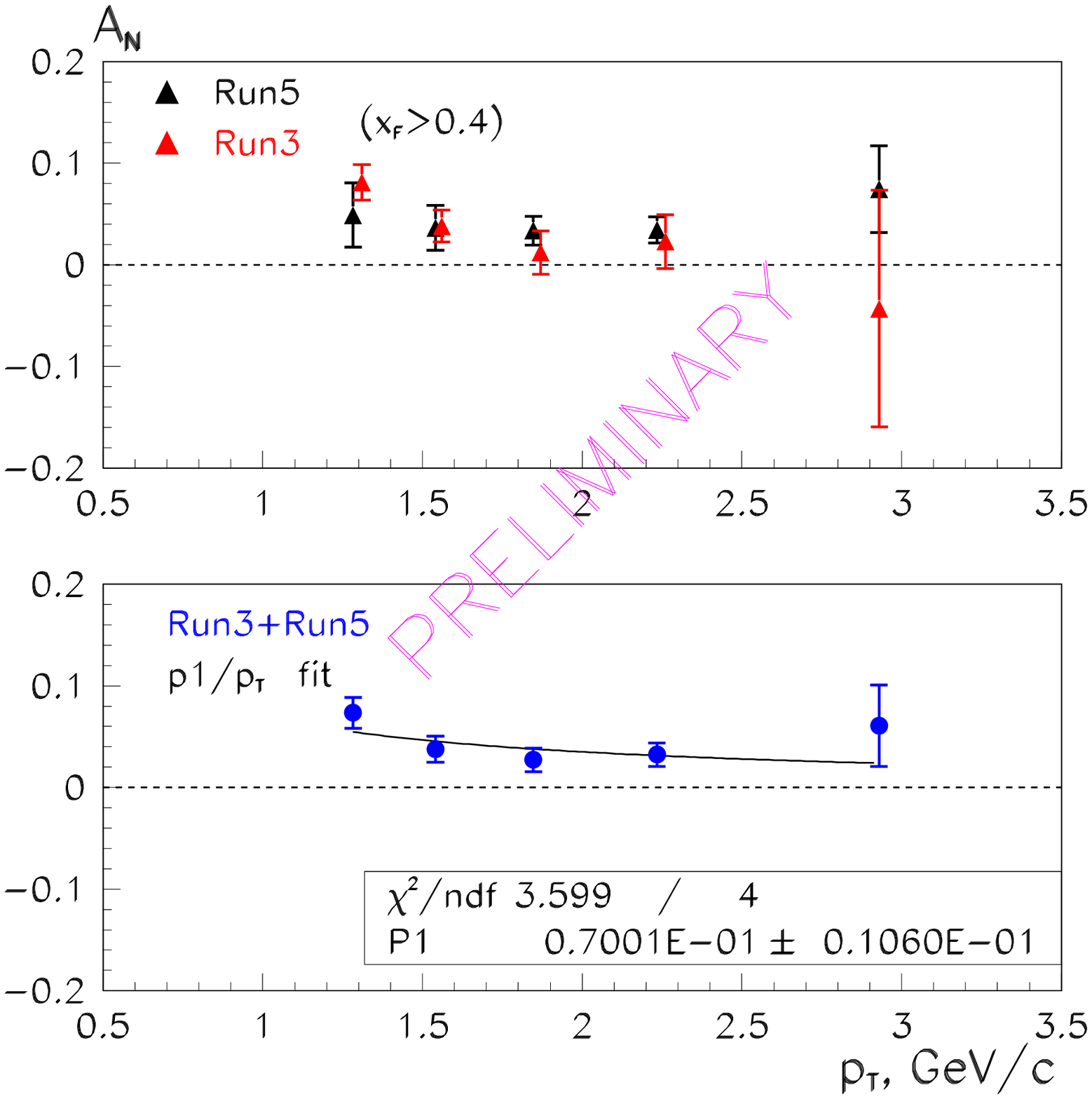}}
  \end{minipage}
\caption{The left panel shows $A_N$ vs.\@ $x_F$ for inclusive $\pi^0$ production at $\langle\eta\rangle$ = 3.7 and 4.0.  The right panels show $A_N$ vs.\@ $p_T$ for those $\pi^0$ with $x_F>0.4$. \label{fig:pi0_AN}}
\end{figure}

Additional $\pi^0$ data were taken using the STAR Forward $\pi^0$ Detector (FPD) during RHIC Runs 3 and 5.  The measured $A_N$ are shown as functions of $x_F$ and $p_T$ in Fig.\@ \ref{fig:pi0_AN}.  For $x_F>0.4$, $A_N$ follows the expected $1/p_T$ dependence.  If all data with $x_F>0.4$ are averaged, $A_N$ is approximately 6 sigma from zero.

Recent calculations\cite{Ansel06} indicate that the maximum possible contribution to the $\pi^0$ $A_N$ from the Collins effect is considerably smaller than that from the quark Sivers effect.  But neither is excluded by the data.  It's very important to isolate the separate contributions from the Sivers and Collins effects.  The former would provide information regarding parton orbital motion, whereas the latter would provide a window to measure transversity.  Gluons can also experience the Sivers effect.  The gluon Sivers effect can make a leading-twist contribution to the back-to-back di-jet angular correlation at mid-rapidity.\cite{Boer04}  The STAR Collaboration transverse spin goals for Run 6 are to elucidate the dynamics underlying the observed $\pi^0$ transverse single-spin asymmetries, and to search for evidence of the gluon Sivers effect in mid-rapidity di-jet production.

\begin{figure}[ht]
  \begin{minipage}{0.64\textwidth}
    {\epsfxsize=\textwidth\epsfbox{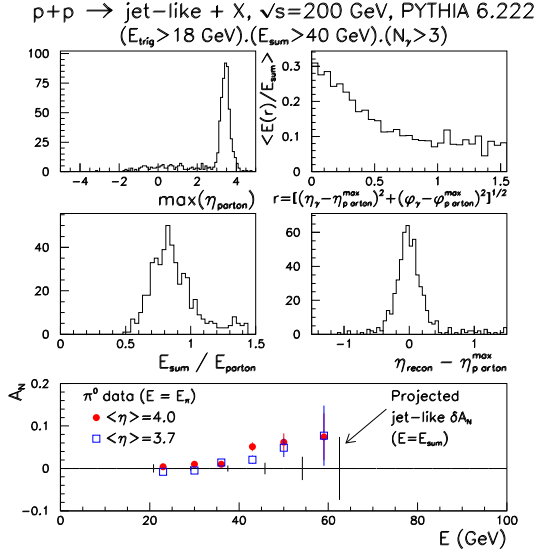}}
  \end{minipage}
  \hfill
  \begin{minipage}{0.34\textwidth}
\caption{PYTHIA simulations of FPD++ response for p+p collisions at $\sqrt{s}=200$ GeV.  Upper left: Pseudorapidity of the most-forward hard-scattered parton.  Upper right:  Distribution of photon energy relative to the thrust axis, showing the jet shape and an underlying event background.  Middle left:  Ratio of the summed photon energy to that of the most forward hard-scattered parton.  Middle right:  Difference between the measured jet pseudorapidity and that of the most forward hard-scattered parton.  Bottom:  Projected statistical precision for jet-like events assuming 5/pb of proton luminosity and 50\% polarization.  \label{fig:jet_sim}}
  \end{minipage}
\end{figure}

\section{Run 6 Measurements}

The most direct way to separate the Sivers and Collins effects is to supplement $\pi^0$ measurements with asymmetries for forward jet and direct photon production.  The Sivers effect introduces a correlation between the incident proton spin and $k_T$ of its constituent partons.  This correlation will also produce transverse single-spin asymmetries for jet and direct photon production.  The final-state jet structure should be azimuthally symmetric.  In contrast, the Collins effect involves a $k_T$ dependence in parton fragmentation.  There should be no asymmetry for complete jets or direct photons.  Rather, the Collins effect implies the existence of an azimuthal asymmetry of the jet structure about the thrust axis.

STAR has expanded the FPD into the FPD++ in order to perform these measurements during Run 6.  Figure \ref{fig:jet_sim} shows PYTHIA simulations of the response of the FPD++ to ``jet-like" events, defined to involve at least 4 detected photons and a large amount of neutral energy.  The simulations\cite{Les_Como} demonstrate that the FPD++ will be able to measure the jet thrust axis and structure reliably.  If the $\pi^0$ $A_N$ arises from the Sivers effect, the FPD++ should observe a 4--5 sigma effect for $A_N$ for jet-like events.  The FPD++ is also large enough to perform the isolation cuts necessary to separate direct photons from $\pi^0$ decays.

\begin{figure}[ht]
    {\epsfxsize=\textwidth\epsfbox{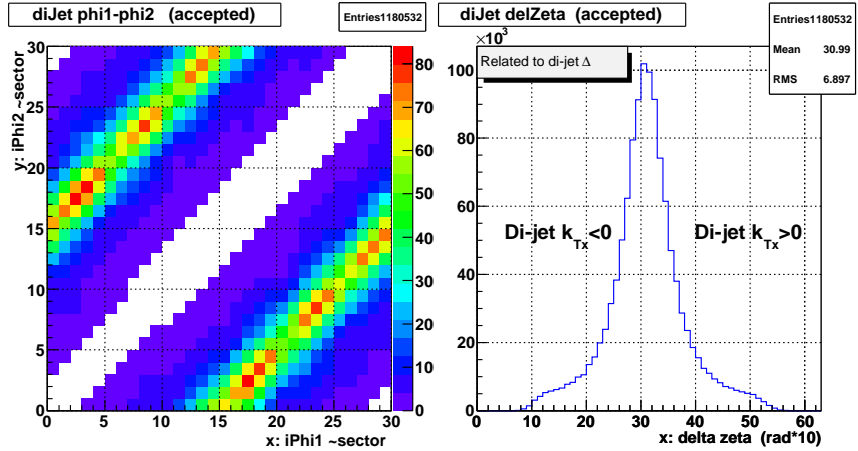}}
\caption{On-line di-jet distributions for accepted events from the level-2 trigger.  Left panel shows $\phi_2$ vs.\@ $\phi_1$.  The six-fold symmetry arises from the ``jet patch" segmentation at level-0.  Right panel shows $\Delta\zeta$, which is related to $\phi_2-\phi_1$ (see text). \label{fig:L2_out}}
\end{figure}

STAR also measures jets at mid-rapidity with the Time Projection Chamber and Barrel and Endcap Electromagnetic Calorimeters (B/EEMC).  For Run 6, a level-2 trigger was implemented to maximize sensitivity to the gluon Sivers effect.\cite{Boer04}  Events containing at least 1 jet at level-0 were selected when two clusters of electromagetic energy separated by at least 60$^{\circ}$ were found.  Figure \ref{fig:L2_out} shows the on-line di-jet angular distributions.  $\Delta\zeta = \pi+|\phi_2-\phi_1-\pi|$ when the di-jet $k_T$ points in the $+x$ direction.  When the di-jet $k_T$ points in the $-x$ direction, $\Delta\zeta = \pi-|\phi_2-\phi_1-\pi|$.  The proton spins are polarized in the $\pm y$ direction, so the gluon Sivers effect will appear as a spin-dependent shift in the $\Delta\zeta$ centroid.

\section{Conclusion}

STAR had a very successful transverse spin run in 2006.  Sufficient data were collected to satisfy both primary goals.  Looking ahead to Run 7, STAR is further expanding the FPD++ into a Forward Meson Spectrometer, which will provide complete electromagnetic calorimetry for $2.5<\eta<4$, and nearly complete coverage over the full range $-1<\eta<4$ when combined with the B/EEMC.


\begin{thebibliography}{0}
\bibitem{pi0_PRL} J. Adams et al. (STAR Collaboration), Phys. Rev. Lett. {\bf 92}, 171801 (2004).
\bibitem{FNAL_pi0} D.L. Adams et al., Phys. Lett. B {\bf 264}, 462 (1991).
\bibitem{Bour04} C. Bourrely and J. Soffer, Eur. Phys. J. C {\bf 36}, 371 (2004).
\bibitem{dAu_PRL} J. Adams et al. (STAR Collaboration), nucl-ex/0602011.
\bibitem{PRL_theory} M. Anselmino et al., Phys. Lett. B {\bf 362}, 164 (1995); M. Anselmino et al., Phys. Rev. D {\bf 60}, 054027 (1999); J. Qiu and G. Sterman, Phys. Rev. D {\bf 59}, 014004 (1998); Y. Koike, AIP Conf. Proc. {\bf 675}, 449 (2003).
\bibitem{DIS06} F. Yuan, these proceedings; P. Mulders, these proceedings.
\bibitem{Ansel06} M. Anselmino et al., Phys. Rev. D {\bf 73}, 014020 (2006).
\bibitem{Boer04} D. Boer and W. Vogelsang, Phys. Rev. D {\bf 69}, 094025 (2004).
\bibitem{Les_Como} L.C. Bland, hep-ex/0602012.

\end{thebibliography}
\end{document}